\documentclass[aps,prd,superscriptaddress,nofootinbib]{revtex4}

\usepackage{float}
\usepackage{amsmath,amssymb}%,graphicx}
\usepackage[dvips]{graphicx}
\DeclareGraphicsExtensions{.jpg,.mps,.pdf,.png,.eps} 
\usepackage{color}

\usepackage[applemac]{inputenc}

\newcommand{\be}{\begin{equation}}
\newcommand{\ee}{\end{equation}}
\newcommand{\bea}{\begin{eqnarray}}
\newcommand{\eea}{\end{eqnarray}}
\newcommand{\non}{\nonumber}

\begin{document}

\title{On the equivalence of Jordan and Einstein frames in scale-invariant gravity}

%%%%%%%%%%%%%%%%%%%  ABSTRACT %%%%%%%%%%%%%%%
\author{Massimiliano Rinaldi}
\email{massimiliano.rinaldi@unitn.it}

\affiliation{Department of Physics, University of Trento\\ Via Sommarive 14, 38123 Trento, Italy}
\affiliation{INFN - TIFPA \\ Via Sommarive 14, 38123 Trento, Italy }

\date{Received: date / Revised version: date}

%%%%%%%%%%%%%%%%%%%  ABSTRACT %%%%%%%%%%%%%%%

\begin{abstract}
\noindent In this note we consider the issue of the classical equivalence of scale-invariant gravity in the Einstein and in the Jordan frames. We first consider the simplest example $f(R)=R^{2}$ and show explicitly that the equivalence breaks down when dealing with Ricci-flat solutions. We discuss the link with the fact that flat solutions in quadratic gravity have zero energy. We also consider the case of  scale-invariant tensor-scalar gravity and general $f(R)$ theories. We argue that all scale-invariant gravity models have Ricci flat solutions in the Jordan frame that cannot be mapped into the Einstein frame. In particular, the Minkowski metric exists only in the Jordan frame. In this sense, the two frames are not equivalent.
\end{abstract}

 \maketitle
 %%%%%%%%%%%%%%%%%%%%%%

 %%%%%%%%%%%%%%%%%%%%%%
 
\section{Introduction}

\noindent The classical physical equivalence between conformally related theories of gravity has been discussed in several papers in the past, sometimes reaching opposite conclusions, see e.g.\ \cite{flana,capo,valerio,deruelle}. Concerning $f(R)$ gravity, the equivalence in the quantum regime was recently discussed in \cite{ruf}, where it was established that one-loop divergences in the two frames are the same, at least on-shell. Previously, this equivalence was proven in the case of constant curvature space in \cite{zerb}.

Past discussions included the precise meaning of ``equivalence''. Here we use a very simple criterium: Einstein and Jordan frames are equivalent if they both have Minkowski space as a solution. Since the lack of Minkowski space is an issue also for perturbative quantum calculations, the non-equivalence automatically expands to the semiclassical level, if not the full quantum one.

In these notes we point out that, for pure quadratic gravity, namely $f(R)=R^{2}$, which is also scale-invariant, the Ricci flat solutions (including Minkowski) are present in the Jordan frame (JF) but are lost in the Einstein frame (EF). At first sight, this is simply a consequence of the fact that the conformal factor is proportional to the curvature and so it vanishes for Ricci-flat spacetimes. However, there might be more. Scale-invariant gravity is known to have Ricci flat solutions whose (ADM-like) energy vanishes  \cite{bhs}. Thus, we can translate what we discussed above by saying that we cannot map JF solutions with zero energy into EF solutions.

We extend our discussion to the scale-invariant $f(R,\phi)$ theory studied in \cite{vanzomax}, where $\phi$ is a real scalar field and we show that the same kind of non-equivalence occurs. Then, we argue that this is a completely general property of any theory that is quadratic in $R$, contains an arbitrary number of fields,  and is scale-invariant. The fundamental reason is that such a generic theory must contain the term $R^{2}$ in the JF. When turned into the EF, this term yields an effective cosmological constant in the action, independently of other scale-invariant operators eventually present in the JF. As a result, the theory in the JF has Ricci flat solutions that cannot exist in the EF.

The plan of the paper is the following. In the next section we recall the fundamental formulae of the simplest model of quadratic gravity in JF end EF. We consider spherically symmetric solutions and in Sec.\ \ref{sec3} and cosmological ones in Sec.\ \ref{sec4} to explicitly show the lack of classical equivalence. We discuss the generic $f(R)$ gravity in Sec.\ \ref{sec5} and we conclude with the most general case in Sec.\ \ref{sec6}.

%%%%%%%%%%%%%%%%%%%%%%

 \section{Pure quadratic gravity in the two frames}\label{sec2}
 
\noindent  Pure quadratic gravity has attracted some interest as it is one-loop renormalisable and ghost-free \cite{lust,lust2}.
The most general action in four dimensions usually involves the square of the Ricci scalar or tensor together with the square of the Weyl tensor. To prove our point it is sufficient consider the minimal version of quadratic gravity, namely the JF Lagrangian
\bea
{\cal L}_{J}={\alpha\over 36}\sqrt{ |g|} R^{2}\,,
\eea
where $\alpha$ is a dimensionless parameter. Let us perform the Weyl transformation 
\bea
\tilde g_{\mu\nu}&=&\Omega^{2} g_{\mu\nu}\,,\\\label{conf}
\Omega&=&{\sqrt{\alpha R}\over 3M}\,,
\eea
where $M$ is an arbitrary mass scale, necessary to keep $\Omega$ dimensionless. 
We find the corresponding Lagrangian in the EF
\bea
{\cal L}_{E}=\sqrt{|\tilde g|}\left[{M^{2}\over 2}(\tilde R-2\Lambda)-\frac12\tilde g^{\mu\nu}\tilde\partial_{\mu}\psi\tilde\partial_{\nu} \psi  \right]\,,
\eea
where
\bea
\Lambda={9M^{2}\over 4\alpha}\,,
\eea
and where the ``scalaron'' field $\psi$ is given by
\bea
\psi=\sqrt{6}M\ln\Omega= \sqrt{3\over 2}M\ln\left(\alpha R\over 9M^{2}\right)\,.
\eea
In the JF, the equation of motion, in the vacuum, reads
\bea\label{eomj}
\square R=0\,.
\eea
In the EF, instead, we have the standard scalar-tensor equations of motion given by
\bea\label{EE}
\tilde G_{\mu\nu}+\Lambda \tilde g_{\mu\nu}=M^{2}\left( \tilde \partial_{\mu}\psi\tilde \partial_{\nu}\psi-\frac12 \tilde g_{\mu\nu}\tilde  \partial_{\mu}\psi\tilde \partial^{\mu}\psi \right)\,,
\eea
where, however, $M$ is technically still arbitrary\footnote{It can be proven that $M$ is a so-called redundant parameter, see \cite{vanzomax} for more details.}. Despite the apparent simplicity of the equations of motion, quadratic gravity has a rich spectrum of interesting solutions in both JF and EF.

 %%%%%%%%%%%%%%%%%%%%%%

\section{Static and spherically symmetric solutions}\label{sec3}

 %%%%%%%%%%%%%%%%%%%%%%

\noindent In this section we analyse some simple solutions to the theories \eqref{eomj} and \eqref{EE} in vacuum.  We first consider black hole solutions and then the cosmological ones. In all cases, these models violate the equivalence, in the sense specified in the introduction, between JF and EF.

\subsection{Black holes in the JF}

\noindent Static and rotating black hole solutions for the theory \eqref{eomj} were studied in \cite{bhr2rot,bhr2stat}. In particular, in \cite{bhr2stat} it was shown that there exists two inequivalent classes of solutions with spherically symmetric and static metric 
\bea
ds^{2}=-N(r)dt^{2}+N^{-1}(r)dr^{2}+r^{2}d\Sigma_{k}^{2}\,,
\eea
where $d\Sigma_{k}^{2}$ is the line element of the two-dimensional horizon space with toroidal ($k=0$), spherical ($k=1$), or hyperbolic topology ($k=-1$).

\noindent {\bf  (anti)-de Sitter-like black holes.} In this case
\bea
N=k-{r_{s}\over r}-{\Lambda r^{2}\over 3}\,,
\eea
where \emph{both} $\Lambda$ and $r_{s}$ are arbitrary integration constants. For these black holes $R=36\Lambda$, which includes also the Ricci-flat case $\Lambda=0$.

\noindent {\bf  Reissner-Nordstr\"om black holes.} In this case, the metric reads
\bea
N=k-{r_{s}\over r}+{r_{q}^{2}\over r^{2}}\,,
\eea
where $r_{s}$ and $r_{q}^{2}$ are again arbitrary integration constants. For this solution, $R=0$ for all parameters.

Among the properties of these solutions, the most peculiar one concerns the entropy. For $f(R)$  gravity, Wald has found that the entropy can be expressed as $S\propto A(df(R)/dR)$, where $A$ is the horizon area. In the quadratic case then $S\propto AR$ and vanishes identically for the second class of solutions, even though $A\neq 0$ and the temperature is well-defined. This seemingly paradoxical result is in line with the fact that these kinds of solutions have zero ADM energy, as proven in \cite{bhs}. Thus the first law of thermodynamics is trivially satisfied (for more details, see the discussion in ref.\ \cite{bhr2stat}).

 %%%%%%%%%%%%%%%%%%%%%%

\subsection{Black holes in the EF}

 %%%%%%%%%%%%%%%%%%%%%%

\noindent In the EF, the theory \eqref{EE} is essentially a tensor-scalar theory with a positive cosmological constant. The scalar field is real, massless, and with a canonical kinetic term. These conditions fulfil  the hypothesis of the no-hair theorem proved in \cite{bhattacharya}, see also \cite{moss}. Such a theorem states that the only static and spherically symmetric solution is the Schwarzschild-(anti) de Sitter with a constant scalar field everywhere. This solution has also positive curvature and strictly positive energy. 

Then, we have an explicit example of the non-equivalence between the solutions space of the two frames: in the JF we have zero energy and Ricci flat solutions (including Minkowski) while in EF we have only positive curvature and energy solutions. Most importantly, there is no Minkowski solution in the EF. This unavoidably raises questions whether a perturbative quantum theory developed around Minkowski space in the JF has an equivalent version in the EF.

% In our specific case it is easy to show that the scalar field must be constant. Let us consider the general time-dependent spherically symmetric  metric
%\bea
%ds^{2}=-N(t,r)dt^{2}+N^{-1}(t,r)dr^{2}+r^{2}d\Omega^{2}\,.
%\eea
%Here, we choose $\Lambda$ positive and, therefore,  the spherical topology for the horizon must be spherical since no black hole solutions are known with different topologies in de Sitter space. 
%We assume that also the scalar field is time-dependent. By summing the $tt$-component and the $rr$-component of Eq.\ \eqref{EE} we find
%\bea
%r{\partial N\over \partial t}+N+r^{2} \Lambda-1=0\,,
%\eea
%which has the general solution
%\bea
%N=1-{c\over r}-{\Lambda r^{2}\over 3}+{g(t)\over r}\,.
%\eea
%Where the quantity $c$ is an arbitrary length and $g(t)$ is an arbitrary function of time. By substituting this back into the $rr$- or $tt$-component of  Eq.\ \eqref{EE} we find the equation for the scalar field
%\bea
%N^{2}\left(\partial\psi\over \partial r\right)^{2}+\left(\partial\psi\over \partial t\right)^{2}=0\,,
%\eea
%which implies that $\psi$ must be constant both in time and space. Finally, the $tr$-component of Eq.\ \eqref{EE} is satisfied only if $g(t)=$ const. Therefore we conclude that the only solution is the Schwarzschild-de Sitter one with constant scalar (real) field. 

 %%%%%%%%%%%%%%%%%%%%%%

\section{Cosmological solutions}\label{sec4}

 %%%%%%%%%%%%%%%%%%%%%%
\noindent The non-equivalence of the two frames persists if one consider cosmological solutions. We now show an explicit example.

\subsection{Cosmological evolution in the JF}

\noindent We adopt the spatially flat metric $ds^{2}=-dt^{2}+a(t)^{2}\delta_{ij}dx^{i}dx^{j}$, where $a$ is the usual scale factor. The equation of motion \eqref{eomj} yields
\bea\label{Heq}
2HH''+H'^{2}+6HH'=0\,,
\eea
where the prime corresponds to the derivative with respect to the conformal time $N=\ln a$. Note that the equation of motion is of second order in $H$. If we consider spatially curved metrics, the equation of motion would be of third order in $a$ and instability issues could arise.

The solution to \eqref{Heq} is
\bea
H=(A+B\,e^{-3N})^{2/3}\,,
\eea
where $A,B$ are integration constant. When $A=0$, the solution can be written as $a=\sqrt{t}$, so it behaves like a radiation-dominated Universe with vanishing Ricci scalar. When $B=0$ the solution has constant $H$ and constant Ricci scalar. By writing the equation of motion as a non-linear dynamical system, one can show that the solution spontaneously flows from the radiation-dominated case with $R=0$ ($N\rightarrow -\infty$) towards the de Sitter case with $R=$ const ($N\rightarrow+\infty$). This simple solution could well be a description of a pre-inflationary era and offers a solution to infrared divergences, see e.g.\ \cite{ruth}.

It is interesting to note that both the Reissner-Nordstr\"om solutions of the previous section and the radiation-dominated Universe solution found here in vacuum are equivalent to standard GR implemented with the electromagnetic tensor. How general this property is for scale-invariant gravity is still unclear and deserves further investigations.

 %%%%%%%%%%%%%%%%%%%%%%

\subsection{Cosmological evolution in the EF}

\noindent In the EF, the relevant equations of motion for a spatially flat metric are the first Friedmann equation (obtained from \eqref{EE}) and the Klein-Gordon one, namely
\bea
\tilde H^{2}={\dot\psi^{2}\over 12\kappa}+{\Lambda\over 3}\,,\quad\quad\ddot\psi+3\tilde H\dot\psi=0\,,
\eea
where $\tilde H=d\ln \tilde a/dt$. The system is invariant under constant shifts of $\psi$, so we define $\phi=\dot \psi$ and we turn the time derivatives into $N-$derivatives as before. Then, the system reduces to
\bea
{\phi'\over \phi}+{\tilde H'\over \tilde H}+3=0\,,\quad \tilde H^{2}\left(1-{\phi^{2}\over 12\kappa}\right)={\Lambda\over 3}\,.
\eea
It is easy then to show that
\bea
\tilde H^{2}={e^{-6(N-N_{0})}\over 12\kappa}+{\Lambda\over 3}\,,
\eea
for arbitrary $N_{0}$, while $\phi$ satisfies the equation 
\bea
\phi \tilde H=e^{-3(N-N_{0})}\,.
\eea
As $N$ grows (i.e. $\tilde a(t)$ increases) the Hubble parameter becomes constant and $\phi$ vanishes (so $\psi$ becomes constant). The solutions asymptotically approaches a de Sitter spacetime, as in the JF. The Ricci scalar reads
\bea
R=12H^{2}+6HH'=4\Lambda-{e^{-6(N-N_{0})}\over 2\kappa}\,,
\eea
and never vanishes (excepts at  critical time that depends on $N_{0}$ and $\Lambda$ so it is arbitrary). Thus no Ricci-flat solutions exist, not even asymptotically as in the JF.

These results clearly show the main point of this work: in the JF and in vacuum, there are classical solutions that have vanishing Ricci scalar. These solutions cannot be mapped to the EF because the conformal factor \eqref{conf} vanishes identically.   In addition, in the EF, Ricci flat solution (including Minkowski space) cannot exist due to the cosmological constant term.   Note  that the cosmological constant $\Lambda=9M^{2}/( 4\alpha)$ is not a perturbative parameter in $\alpha$. In addition, the parameter $M$ in the EF should be identified with the Planck mass and cannot be taken arbitrarily small.

 %%%%%%%%%%%%%%%%%%%%%%

\section{The $f(R)$ case}\label{sec5}

\noindent We now discuss under what circumstances the results found above hold for $f(R)$ gravity. In general, $f(R)$ theories are not scale-invariant so the vacuum Lagrangian reads 
\bea
{\cal L}_{J}={1\over 2\kappa^{2}}\sqrt{|g|}\,f(R)\,,
\eea
where $\kappa^{2}=8\pi G$. Under the conformal transformation $\tilde g_{\mu\nu}=\Omega^{2}g_{\mu\nu}$, with 
\bea\label{conff}
\Omega^{2}=f'(R)\,,
\eea
(where the prime now indicates a derivative with respect to $R$) we obtain the Lagrangian in the EF (see \cite{defelice})
\bea
{\cal L}_{E}=\sqrt{|g|}\left[ {1\over 2\kappa^{2}}\tilde R-\frac12\tilde g^{\mu\nu}\tilde\partial_{\mu}\psi\tilde\partial_{\nu} \psi-V(\psi) \right]\,,
\eea
where, as before, $\kappa\psi=2\sqrt{3/2}\ln \Omega$. Unlike the quadratic case, the potential is not constant and reads
\bea\label{pot}
V(\psi)={{Rf'-f}\over 2\kappa^{2}f'^{2}}\,.
\eea
We see that, in general, a vanishing conformal factor implies a diverging potential thus the theory is meaningless in the EF. The only exceptions are  $f(R)=R^{2}$, for which $V$ is constant, and $f(R)=R$ for which $V=0$ and the theory is equivalent to GR.

There are several examples of $f(R)$ theories that feature spherically symmetric solutions, and such that $f'(0)=f(0)=0$, see \cite{calza}. Typically, these models do not have a GR limit in the low curvature regime, namely $\lim_{R\rightarrow 0}f(R)\neq R$ so the potential \eqref{pot} does not vanish in this limit and Ricci flat solutions in the EF are not allowed. 

These examples clearly shows that the non-equivalence between frames goes beyond the set of scale-invariant gravity. However, below we argue that if a theory is scale-invariant then JF and EF are not equivalent.

\section{Discussion}\label{sec6}

\noindent We have seen that in quadratic and scale-invariant gravity the solution space of the  Jordan frame cannot be entirely mapped into the solution space of the Einstein frame. This is more than a mere mathematical problem (aka the vanishing of the conformal factor). In fact, this can be linked also to the energy that we associate to the solution. Indeed, according to the theorem of Boulware et al. \cite{bhs}  (see also \cite{deser1,deser2}), asymptotically flat solutions to the $R^{2}$ theory  have zero ADM energy \footnote{See however \cite{deser3}, where an alternative definition of energy is discussed.}. Since the ADM energy is not a conformally invariant quantity, it is to be expected that its value changes upon conformal transformations which, in order to be smooth, should map a zero energy solution of the JF to an equivalent zero energy solution in the EF. But such solutions do not exist, as explicitly shown in the case of spherically symmetric solutions that brings along an unavoidable strictly positive cosmological constant.

The question now is whether this conclusion can be extended to any scale-invariant theory. In other words, can we claim that all scale-invariant gravitational theories in JF are not equivalent to their EF counterparts? For sure, this statement holds for the extension of the $f(R)=R^2$ model obtained by adding all scale-invariant operators involving curvature and a scalar field, studied in \cite{vanzomax}. The action reads
\bea\label{phiac}
{\cal L}_{J}=\sqrt{-\det g}\left[{\alpha\over 36}R^{2}+{\xi\over 6}\phi^{2}R-\frac12(\partial \phi)^{2}-{\lambda\over 4}\phi^{4}\right]\,.
\eea
In the EF, the action becomes
\bea
{\cal L}_{E}&=&\sqrt{\tilde g}\Bigg[ {M^{2}\over 2}\tilde R-\frac12\tilde g^{\mu\nu}\partial_{\mu}\psi\partial_{\nu}\psi-\frac12\exp\left(-{\sqrt{2}\psi\over\sqrt{3}M}\right)\tilde g^{\mu\nu}\partial_{\mu}\phi\partial_{\nu}\phi\\\non
&-&V(\phi,\psi) -{9 M^{4}\over 4\alpha}\Bigg],
\eea
where
\bea
V(\phi,\psi)={\lambda\phi^{4}\over 2}\exp\left(-{2\sqrt{2}\psi\over\sqrt{3}M}\right)-{3\lambda M^{2}\phi^{2}\over 2\xi}\exp\left(-{\sqrt{2}\psi\over\sqrt{3}M}\right)\,.
\eea
The conformal factor is given by
\bea\label{CF}
\Omega^{2}={1\over M^{2}}\left({\xi \phi^{2}\over 3}+{\alpha R\over 9}\right)\,.
\eea
We note that the cosmological constant is still present in the EF action. In fact, the latter appears essentially from the conformal transformation of the term $R^{2}$ alone and it persists when we add other scale-invariant operators \footnote{In \cite{vanzomax} it appears that a vanishing $\lambda$ sets to zero the cosmological constant. However, in that paper the authors have initially set $\alpha=\xi^{2}/\lambda$ so this option is actually not feasible.}. Again, we have an action that does not allow Ricci flat solutions in the EF, while in the JF these are present (it is sufficient to set $\phi=0$ in \eqref{phiac}). Also in this case, Ricci flat solutions in the JF implies the vanishing of the conformal factor \eqref{CF}. In fact, to be flat the solutions needs not only $R=0$ but also $\phi=0$, otherwise we would end up with an effective cosmological constant in \eqref{phiac}.

In summary, we conclude that all gravitational theories that are scale-invariant and contains the $R^{2}$ term have the same problem, namely that they have Ricci flat solutions that cannot be mapped to the EF because the conformal factor vanishes. Most likely, this is linked to the fact that flat and scale-invariant solutions have zero energy and these cannot be mapped into positive energy field configurations. In particular, while the JF admits Minkowski space as a solution, the EF does not. Thus, the expansion of quantum operators around flat space is perfectly defined in the JF but not in the EF. At these levels, the two frames are certainly not equivalent.

\section*{Acknowledgments}

\noindent Part of this work was realised during my stay at the Department of Physics and Astronomy of the McMaster University and at the Perimeter Institute for Theoretical Physics, which I wish to thank for kind hospitality. Many thanks to prof.\ C.\ Burgess and his group for insightful discussions, and to V.\ Faraoni, L.\ Sebastiani, L.\ Vanzo, and S.\ Zerbini for helpful suggestions.

 %%%%%%%%%%%%%%%%%%%%%%
 %%%%%%%%%%%%%%%%%%%%%%

\end{document}